\documentclass[aps,preprint,showpacs]{revtex4}
\usepackage{amsmath}
\usepackage{graphicx}

\begin{document}

\title{Plasmonic atoms and plasmonic molecules}

\author{D.V. Guzatov and V.V. Klimov}
\email{vklim@sci.lebedev.ru}
 \affiliation{P.N. Lebedev Physical
Institute, Russian Academy of Sciences, 53 Leninsky Prospect,
Moscow 119991, Russia}

\begin{abstract}
The proposed paradigm of plasmonic atoms and plasmonic molecules
allows one to describe and predict  the strongly localized
plasmonic oscillations in the clusters of nanoparticles and some
other nanostructures in uniform way. Strongly localized plasmonic
molecules near the contacting surfaces might become the
fundamental elements (by analogy with Lego bricks) for a
construction of fully integrated opto-electronic nanodevices of
any complexity and scale of integration.
\end{abstract}

\date{\today}
 \maketitle

\section{Introduction}

\label{intro}

A technological break-through in the fabrication of nanodimensional clusters
and other metallic nanoparticles gave rise to a development of such
nano-technological and nanooptical branch as nanoplasmonics, which is of
great interest to physicists, chemists, material engineers, IT specialists,
and biologists. The nanoplasmonics deals with conduction electron liquid
oscillations in metallic nanostructures and nanoparticles, and an
interaction of those oscillations with light (plasmon-polariton) \cite%
{ref1,ref2}. Such studies are also aimed at the oscillations of a crystal
lattice (SiC, for example) in the nanoparticles, whose interaction with
light has much in common with the plasmon oscillations (phonon-polariton)
\cite{ref3}.

The important feature of nanoplasmonic phenomena is the combination of a
strong spatial localization and high-frequency (from ultraviolet to
infrared) of electron oscillations. In its turn, strong localization leads
to a giant enhancement of the local optical and electrical fields. These
important features of plasmonic particles made it possible to discover quite
a number of new effects. One of the most developed is the use of large local
fields near plasmonic nanoparticles for enhancement of the Raman scattering
cross-section. Recent experiments have shown that such an increase may
achieve 10-14 orders of magnitude, which may help to resolve single
molecules \cite{ref4}-\cite{ref7}. The local enhancement of the fields can
also be used to increase the fluorescence intensity and to determine the
structure of a single DNA strand without using the fluorescent labels \cite%
{ref8}-\cite{ref9}. By using the nanoparticles of more complex shape one can
provide enhancement of both the absorption and the emission of light by
natural and artificial fluorophores \cite{ref10}. On the other hand, the
plasmon nanoparticles are proposed to be used in nanolasers \cite{ref11} and
to stimulate plasmonic oscillations in nanoparticles by means of the optical
emission (SPASER) \cite{ref12}. Beside these new applications of the
plasmonic nanoparticles, one can essentially increase the efficiency-cost
ratio, for example, in solar batteries or light emitting diodes \cite%
{ref13,ref14} by using the achievements in nanoplasmonics. And finally, it
is awaited that the nanoplasmonics will make it possible to create a new
element base (Genuine Integrated Optics and Ultracompact Optical Components)
for the computers and data processing equipment by taking an advantage of
small dimensions of metallic nanoparticles and fast speed of optical
processes \cite{ref15}.

An intricate spatial structure of the physical phenomena, which form the
basis of nanoplasmonics, impedes a development of the latter. Very often,
the numerical studies do not allow one to explain the physics of the
observed phenomena, while the analytical studies are mostly devoted to the
case of spherical and spheroidal nanoparticles, which are very far from the
synthesized nanostructures from the viewpoint of geometry and physics.
Separate nanoparticles are often approximated by point dipoles. This permits
one to explain several phenomena (for example, a propagation of plasmons in
a chain of nanoparticles \cite{ref16}). But as a whole, the physics of
plasmonic oscillations in complex nanostructures and nanoparticles remains
insufficiently studied, and to make progress in this field one needs new
approaches and new ideas.

As one of such approaches we suggest that the plasmonic effects observed in
clusters of nanoparticles should be described by the paradigm of plasmonic
atoms and molecules, which has much in common with the paradigm of normal
atoms and molecules. This would allow one to explain the complex phenomena
of nanoplasmonics in terms of atomic and molecular physics, and to propose
quite a number of new applications of the nanoplasmonics.

Within the framework of our approach we propose to treat the
plasmonic oscillations of the connected nanoparticles as the
plasmonic atoms. The plasmonic atoms are analogous, in a certain
way, to the electron oscillations at the nanocrystal quantum dots,
which are also often called the artificial atoms \cite{ref17}.

The bound states of plasmonic atoms in the clusters of
nanoparticles are proposed to be called the plasmonic molecules,
and finally, the plasmonic oscillations in the nanoparticle
gratings are proposed to be called the plasmonic crystals. Such an
approach seems to be efficient and useful as it follows from the
strongly localized plasmonic oscillations recently discovered in
the clusters of spherical nanoparticles \cite{ref18,ref19,ref19a}.
These oscillations, being localized and of short-range, represent
the plasmonic molecules, and on their basis it is possible to
create different kinds of plasmonic nanostructures and to develop
new nanodevices (optical nanosensors of single molecules,
nanoelectromechanical systems (NEMS \cite{ref20}), and even
plasmonic computers).

\section{Plasmonic atoms}

\label{sec:1}

Let us consider in more detail the analogy between the plasmonic and
ordinary atoms. In the simplest case of a hydrogen atom, the energy levels
are defined by the Bohr equation

\begin{equation}
E_{n}=-\frac{E_{1}}{n^{2}},\quad n=1,\;2,\;3,\;\ldots  \label{eq1}
\end{equation}

\noindent where $E_{1} $=13.55 eV is the ionization energy of hydrogen atom.
The wave function of an electron has the form, in this case:

\begin{equation}
\psi _{nlm}\left( r,\theta ,\varphi \right) =Z_{nl}\left( r\right)
Y_{l}^{m}\left( \theta ,\varphi \right) ,  \label{eq2}
\end{equation}

\noindent where $Z_{nl}$ is the radial part of the wave function, and $%
Y_{l}^{m}$, the spherical harmonic. For example, in the case of the first
excited (2P) atomic state, the expression for the radial function takes the
form:

\begin{equation}
Z_{21}\left( r\right) =N_{21}\left( \frac{r}{r_{B}}\right) e^{-\frac{r}{%
2r_{B}}},  \label{eq3}
\end{equation}

\noindent where $N_{21} $ is the normalized constant, and $r_{B} $, the
radius of the first Bohr orbit. Function (\ref{eq3}) is represented in Fig.~%
\ref{fig:1}.

In the case of a spherically symmetric plasmonic atom, i.e. the plasmonic
oscillations in a metallic spherical nanoparticle of the radius $R_{0}$,
under the Drude dispersion law $\varepsilon \left( \omega \right) =1-\left(
\omega _{pl}/\omega \right) ^{2}$ (where $\omega _{pl}$ is bulk plasmon
frequency), the spectrum of the plasmonic oscillations will take the form:

\begin{equation}
E_{n}=\hbar \omega _{pl}\sqrt{\frac{n}{2n+1}},\quad n=1,\;2,\;3,\;\ldots
\label{eq4}
\end{equation}

\noindent whereas the electric field potential, in respect to (\ref{eq4}),
will have the form corresponding to (\ref{eq2}) with the radial wave function

\begin{equation}
Z_{n}=N_{n}\left\{
\begin{array}{cc}
\left( r/R_{0}\right) ^{n}, & r\leq R_{0}, \\
\left( R_{0}/r\right) ^{n+1}, & r>R_{0},%
\end{array}%
\right.  \label{eq5}
\end{equation}

\noindent where $n=1,2,3,...$ and $N_{n}$ is some  constant. Function (\ref%
{eq5}) for the index $n=1$ is illustrated in Fig.~\ref{fig:1} by a dotted
line. It is seen from this figure that wave functions of the ordinary and
plasmonic atoms are similar. The spectra (\ref{eq1}) and (\ref{eq4}) also
have much in common.

The plasmonic oscillations have similar properties in the other spherically
symmetrical layered systems (nanomatreshka) \cite{ref21} and spheroidal
nanoparticles (nanorice) \cite{ref22}.

The resonant plasmonic frequencies and potentials of plasmonic
atoms in more complex and less symmetrical nanoparticles have a
more complex form. For example, in the case of a metallic
nanoparticle having a form of a three-axial ellipsoid with the
semi-axes $a_{1}>a_{2}>a_{3}$, the plasmonic frequencies will be
defined by the ratio \cite{ref24}

\begin{eqnarray}
\omega _{nm} &=&\frac{\omega _{pl}}{\sqrt{1-\varepsilon _{nm}}},\quad
\varepsilon _{nm}=\frac{E_{n}^{m}\left( a_{1}\right) F_{n}^{\prime
}{}^{m}\left( a_{1}\right) }{E_{n}^{\prime }{}^{m}\left( a_{1}\right)
F_{n}^{m}\left( a_{1}\right) },  \notag \\
n &=&1,\;2,\;3,\;\ldots \quad m=1,\;2,\;...\;2n+1,  \label{eq6}
\end{eqnarray}

\noindent where $E_{n}^{m}$ and $F_{n}^{m}$ are the internal and external
Lame functions \cite{ref25}, and a stroke denotes the derivative of function
by its argument. The dependences of the resonant dielectric permeabilities
for $n=3$ are shown in Fig.~\ref{fig:2} . The explicit
expressions for eigen frequencies and eigen functions, e.g., for the case of
$n=3$ and $m=7$, have the form (Guzatov, Klimov to be published)

\begin{eqnarray}
\omega _{37} &=&\omega _{pl}\sqrt{\frac{\left( a_{1}a_{2}a_{3}\right) ^{3}}{2%
}\left( \sum\limits_{\alpha =1}^{3}a_{\alpha }^{-2}\right) I_{123}},  \notag
\\
\psi _{37} &=&C_{37}xyz,  \label{eq7}
\end{eqnarray}

\noindent where $C_{37}$ is a constant ; and
$I_{123}=\int\limits_{0}^{\infty }\frac{du}{\left\{ \left(
u+a_{1}^{2}\right) \left( u+a_{2}^{2}\right) \left(
u+a_{3}^{2}\right) \right\} ^{3/2}}$.

The distribution of a surface charge for the plasmonic oscillations with $%
n=3 $ and $m=7$ derived from Eq. (\ref{eq7}) has the form depicted in Fig.~%
\ref{fig:3} (Guzatov, Klimov to be published).

Even more complex properties characterize the cubic and polyhedral plasmonic
atoms \cite{ref26}-\cite{ref31}. The spectrum of the cubic plasmonic atoms was
studied in \cite{ref26}, and some of the first values take the form

\begin{equation}
\frac{E_{n}}{\hbar \omega _{pl}}=0.46225,\quad 0.54473,\quad 0.58722,\quad
0.66372,\quad 0.74953,\quad 0.83918  \label{eq8}
\end{equation}

Figure ~\ref{fig:4} illustrates schematically the surface charge
distribution for several states of a cubic plasmonic atom as found in \cite%
{ref27}. The experiments studying the cubic plasmonic atoms in \cite{ref31}
proved to be in an agreement with the theoretical calculations.

It is very important that both the plasmonic atom wavefunction (electric
potential) and the field intensity are changing within the whole volume of a
nanoparticle and the adjacent space, despite the fact that the structure of
plasmonic atoms may be rather complex. By analogy with that, the surface
charge in any plasmonic atom changes noticeably over the whole nanoparticle
surface. Such a smeared spatial structure of plasmonic atoms is their
characteristic feature, and can be used in quite a number of applications
such as nanosensors of single molecules. However, as the distance between
nanoparticles decreases they start to interact more intensively. And instead of
the plasmonic atoms it is expedient to use the conception of plasmonic
molecules to describe the plasmonic oscillations in the closely situated and
strongly interacting nanoparticles. As in the case of ordinary atoms and
molecules, the question about the existence of bound states of the plasmonic
atoms emerges here first of all.

\section{Plasmonic molecules in a two-sphere cluster}

\label{sec:2}

The evidence for the existence of bound states of the plasmonic atoms was
first provided in \cite{ref18,ref19} by the example of a cluster of two
closely sitiated spherical nanoparticles (Fig.~\ref{fig:5}).

Generally speaking, the plasmonic oscillations in the cluster of two
nanoparticles were studied earlier too \cite{ref32}-\cite{ref37}. However,
it was assumed, both explicitly and implicitly, that the bound states are
not formed in such a cluster, and the interaction assumes a scattering of
plasmonic atoms on each other with a respective splitting of energy levels,
and hybridization of wave functions (potentials) of single plasmonic atoms
\cite{ref36,ref37}. That was the reason why the plasmonic molecules had not
been predicted earlier.

The discovered by us spectrum of the plasmonic oscillations in a cluster of
two  spherical nanoparticles is shown in Fig.~\ref{fig:6} for the case of
identical spheres ($R_{1}=R_{2}=R_{0})$. There are only symmetrical
(T-modes)  and antisymmetrical (L-modes) states of the scattering within the
range of $\omega <\omega _{pl}/\sqrt{2}$ (${\varepsilon <-1}$). They exist
at any distance between the particles, and continuously transit into the
respective states of the weakly interacting plasmonic atoms with the
characteristics (\ref{eq4}) and (\ref{eq5}), within large distances between
the nanospheres. Just those scattering states had become the subject of
investigations in \cite{ref32}-\cite{ref37}.

In the region of $\omega _{pl}\geq \omega >\omega _{pl}/\sqrt{2}$ (${%
0>\varepsilon >-1}$) the plasmonic oscillations may exist only at small
distances between the nanoparticles, and correspond to the bound states of
the plasmonic atoms and molecules (M-modes). Note that the existence of the
plasmonic molecules is by no means connected with a quasistatical
approximation, and the calculations made within the framework of the total
system of Maxwell's equations, with account for retardation effects, provide
evidence for the existence of plasmonic molecules in the case of a small gap
as compared with the radii of spheres (Guzatov, Klimov unpublished).

The analogous spectra appear for higher azimuthal numbers $m$, although
there are some peculiarities in the spectra of the unbound plasmonic atoms.

In the region of small distances between the nanospheres, the properties of
the plasmonic molecules may be studied analytically within quasistatic
approximation. For the Drude dispersion law, the spectrum of the plasmonic
molecules takes the form:

\begin{eqnarray}
\omega _{m}^{M} &=&\frac{\omega _{pl}}{\sqrt{1-\varepsilon _{m}^{M}}},
\notag \\
\varepsilon _{m}^{M} &=&-\left( M+m+\delta _{m}\right) \text{acosh}\left(
R_{12}/\left( 2R_{0}\right) \right) +\ldots  \notag \\
M &=&1,\;2,\;3,\;\ldots \quad m=0,\;1,\;2\;\ldots  \label{eq9}
\end{eqnarray}

\noindent where $R_{12}$ is the distance between the  centers of spheres and

\begin{eqnarray}
\delta _{0} &=&1/2,\quad \delta _{1}=-0.08578,\quad \delta
_{2}=-0.2639,\;...\;\delta _{\infty }=-1/2,  \notag \\
\delta _{m} &=&-\frac{{1}}{{2}}-\frac{{1}}{2m}+\frac{{1}}{8m^{3}}-\frac{{1}}{%
16m^{5}}+\frac{{5}}{128m^{7}}-\ldots \quad \left( m>0\right)  \label{eq10}
\end{eqnarray}

The potential of the axis-symmetric ($m$=0) state of the plasmonic molecule
in a space between the spheres ($-\eta _{0}<\eta <\eta _{0})$ has the form:

\begin{eqnarray}
\psi _{M,m=0} &\approx &\frac{\sqrt{\cosh \eta -\cos \xi }}{a}\left( \frac{{1%
}}{M}\sum\limits_{n=0}^{M-1}\frac{\cosh \left( \left( n+1/2\right) \eta
\right) }{{\cosh }\left( \left( n+1/2\right) \eta _{0}\right) }P_{n}\left(
\cos \xi \right) \right.  \notag \\
&&\left. -\frac{{\cosh }\left( \left( M+1/2\right) \eta \right) }{{\cosh }%
\left( \left( M+1/2\right) \eta _{0}\right) }P_{M}\left( \cos \xi \right)
\right) ,  \label{eq11}
\end{eqnarray}

\noindent where $P_{n}^{m}$ is the associated Legendre function and $\eta
,\;\xi $, the bispherical coordinates

\begin{eqnarray}
\coth \eta  &=&\frac{x^{2}+y^{2}+z^{2}+a^{2}}{2az},\quad \cot \xi =\frac{%
x^{2}+y^{2}+z^{2}-a^{2}}{2a\sqrt{x^{2}+y^{2}}},  \notag \\ a
&=&\sqrt{R_{12}^{2}/4-R_{0}^{2}},  \label{eq12}
\end{eqnarray}

\noindent and where $\cosh \eta _{0}=R_{12}/\left( 2R_{0}\right) $.

For the antisymmetric unbound plasmonic atom we have respectively

\begin{eqnarray}
\omega _{m}^{L} &=&\frac{\omega _{pl}}{\sqrt{1-\varepsilon _{m}^{L}}},
\notag \\
\varepsilon _{m}^{L} &=&-\left( L+m-1/2\right) ^{-1}/\text{acosh}\left(
R_{12}/\left( 2R_{0}\right) \right) +\ldots  \notag \\
L &=&1,\;2,\;3,\;\ldots \quad m=0,\;1,\;2\;\ldots  \label{eq13}
\end{eqnarray}

\noindent and

\begin{eqnarray}
\psi _{Lm} &\approx &\sqrt{\cosh \eta -\cos \xi }  \notag \\
&\times &\frac{{\sinh }\left( \left( L+m-1/2\right) \eta \right) }{{\sinh }%
\left( \left( L+m-1/2\right) \eta _{0}\right) }P_{L+m-1}^{m}\left( \cos \xi
\right) \cos \left( m\varphi \right) .  \label{eq14}
\end{eqnarray}

In the case of large azimuthal numbers, $m\gg 1$, one can also find simple
asymptotic equations for the spectra of plasmonic oscillations in a cluster
of two spheres. This case is of importance for calculation of van der Waals
forces between nanoparticles.

Consider in more detail a structure of the wave functions (electric
potentials) of the plasmonic molecules. Figure ~\ref{fig:7} illustrates a
wave function spatial distribution of a plasmonic molecule and unbound
states of plasmonic atoms in the $x$-$z$ plane.

Spatial structure of antisymmetrical (L-modes) and symmetrical (T-modes)
wave functions of the unbound plasmonic atoms, in the axis-symmetrical case (%
$m=0$) corresponds to the structure of wave functions of the isolated atoms.
Namely, a positive charge lies in one part of a sphere, while a negative
charge, which is equal to the positive one due to the electrical neutrality
of spheres, lies in the opposite part of the latter. An interaction between
the plasmonic atoms is reduced, in this case, to a charge re-distribution on
opposite semi-spheres.

In the case of plasmonic molecules (symmetrical M-modes), the situation is
the opposite, and both positive and negative charges are concentrated in a
small region near a gap between the nanospheres. Far from the gap, the wave
functions of plasmonic molecules practically vanish.

In the case of $m=1$, where the wave functions have an angular dependence $%
\cos \varphi $ or $\sin \varphi $, the situation is analogous. The only
difference is that the dipole moments of the unbound plasmonic atoms and
molecules are directed along the $x$- or $y$-axis.

As a whole, the wave function of the plasmonic atoms remains more or less
distributed over the volume of both nanospheres, whereas the plasmonic
molecules are strongly localized near the gap between the nanospheres in a
region with a characteristic dimension of the order of the gap size with the
maxima at spherical surfaces.

A strong localization of the plasmonic modes is also well seen in Fig.~\ref%
{fig:8}, which shows a distribution of the (surface) charge over a  surface
of spheres. To plot these distributions we have used a polar system of
coordinates ($\rho $, $\theta $) for each sphere in plane $y=0$. The surface
charge distribution was depicted with $\rho =R_{0}\left( 1+\tilde{\sigma}%
\left( \theta \right) \right) $ polar curve where $\tilde{\sigma}\left(
\theta \right) $ is the dimensionless surface charge density as a function
of polar angle of corresponding sphere.

As the distance between the spheres increases, the localization of the
plasmonic molecules decreases. At critical distance between the spheres the
plasmonic molecules disappear, whereas the unbound atoms do not change.

The localization of plasmonic molecules and unbound plasmonic atoms is
different, which defines their different dependence on the exciting
radiation. The polarizability of the unbound plasmonic atoms is of the order
of a nanosphere volume $\alpha \sim R_{0}^{3}$, and they effectively
interact with the uniform external fields of the respective orientation and
symmetry. And on the contrary, the plasmonic molecules have a comparatively
small polarizability $\alpha \sim \Delta ^{3}$, where $\Delta $ is the value
of the gap between the spheres. As a result, the plasmonic molecules are
weakly excited by the uniform optical fields (as compared to the unbound
plasmonic states). On the other hand, the plasmonic molecules interact
effectively with strongly inhomogeneous fields that are localized near the
gap between the spheres. Such fields are produced by the atomic and
molecular radiation near the gap. This makes the plasmonic molecules
extremely perspective from the viewpoint of a development of nanosensors and
elements of the nanodevices, which are sensitive to single molecules. Quite
probably, that the SERS from single molecules \cite{ref4}-\cite{ref7} is due
to their interaction with the plasmonic molecules. It seems that hot points
in fractal structures (Shalaev's hot spots \cite{ref39}) are also due to the
plasmonic molecules.

\section{Plasmonic molecules in more complicated systems of nanoparticles}

\label{sec:3}

Above we considered the plasmonic molecules and unbound states of the
plasmonic atoms formed in a cluster of two identical nanospheres. However,
the plasmonic molecules may be formed within a much wider set of clusters of
the nanoparticles. The only necessary condition for the formation of
plasmonic molecules is smallness of the gap between the nanoparticles of the
finite volume as compared to their characteristic dimensions. The plasmonic
molecules may also be formed in connected particles with two or more
nanobubbles or dielectric nanospheres inside them (Fig.~\ref{fig:9}).

\subsection{Cluster made from different nanospheres}

\label{sec:31}

First of all, the plasmonic molecules can exist in a cluster of two
different nanospheres. Here the spectra and wave functions of unbound
plasmonic atoms and plasmonic molecules are completely analogous to the case
of identical nanospheres. If the distance between the spheres is small, then
the plasmonic molecule spectrum may be written in the form:

\begin{eqnarray}
&&\left( \frac{\varepsilon _{1}\left( \omega \right) -\varepsilon _{3}}{%
\varepsilon _{1}\left( \omega \right) +\varepsilon _{3}}\right) \left( \frac{%
\varepsilon _{2}\left( \omega \right) -\varepsilon _{3}}{\varepsilon
_{2}\left( \omega \right) +\varepsilon _{3}}\right)   \notag \\
&\approx &\exp \left[ \left( 2N+2m-1\right) \Omega \right] ,  \notag \\
N &=&1,\;2,\;3,\;\ldots \quad m\gg 1,  \label{eq15}
\end{eqnarray}

\noindent where the parameter $\Omega $ can be found from the expression

\begin{equation}
\cosh \Omega =\frac{R_{12}^{2}-R_{1}^{2}-R_{2}^{2}}{2R_{1}R_{2}},
\label{eq16}
\end{equation}

\noindent and where $\varepsilon _{1},\;\varepsilon _{2},\;\varepsilon _{3}$
are the dielectric constants of the first and second spheres and the space
between them, correspondingly.

For the dispersion law of Drude for both spheres it can be found from (\ref%
{eq15})

\begin{eqnarray}
\omega _{Nm}^{2} &=&\frac{1}{4}\left( \omega _{pl,1}^{2}+\omega
_{pl,2}^{2}\right) \pm \frac{1}{4}\left\{ \left( \omega _{pl,1}^{2}-\omega
_{pl,2}^{2}\right) ^{2}\right.   \notag \\
&&\left. +4\omega _{pl,1}^{2}\omega _{pl,2}^{2}\exp \left[ -\left(
2N+2m-1\right) \Omega \right] \right\} ^{1/2},
\label{eq17}
\end{eqnarray}

\noindent where $\omega _{pl,1} $ and $\omega _{pl,2} $ are the bulk plasmon
frequencies of spheres.

\subsection{Cluster made of two nonspherical particles}

\label{sec:32}

Strong localization of plasmonic molecules allows one to suggest that their
characteristics depend mainly on the radius of curvature of almost
contacting surfaces and the distance between the centers of the inscribed
(or circumscribed) spheres. So, for any smooth nonspherical particle
the properties of plasmonic molecules  formed in the gap may be estimated by
considering the plasmonic molecules in two spheres, which approximate the
nonspherical particles in the contact point (see Fig.~\ref{fig:10}).

For the closely set semi-infinite bodies the situation turns to be more
intricate, since in this case part of the charges may go arbitrary far from
the contact area. Nevertheless, here exist a localized plasmonic
oscillations, which are analogous to the antisymmetric unbound plasmonic atoms
(L-modes) and plasmonic molecules (M-modes).

In case of closely set semi-infinite hyperboloids
(Fig.~\ref{fig:11}a) the variables in Laplace equation can be
separated within prolate spheroid coordinates \cite{ref40}, and
the localized plasmonic oscillation spectra have the form

\begin{eqnarray}
\omega _{m}^{L} &=&\omega _{pl}\left\{ P_{-1/2+iL}^{\prime }{}^{m}\left( \mu
_{0}\right) \left( P_{-1/2+iL}^{m}\left( -\mu _{0}\right)
-P_{-1/2+iL}^{m}\left( \mu _{0}\right) \right) \right.   \notag \\
&&\times \left( P_{-1/2+iL}^{\prime }{}^{m}\left( \mu _{0}\right)
P_{-1/2+iL}^{m}\left( -\mu _{0}\right) \right.   \notag \\
&&\left. \left. +P_{-1/2+iL}^{\prime }{}^{m}\left( -\mu _{0}\right)
P_{-1/2+iL}^{m}\left( \mu _{0}\right) \right) ^{-1}\right\} ^{1/2},  \notag
\\
\omega _{m}^{M} &=&\omega _{pl}\left\{ P_{-1/2+iM}^{\prime }{}^{m}\left( \mu
_{0}\right) \left( P_{-1/2+iM}^{m}\left( -\mu _{0}\right)
+P_{-1/2+iM}^{m}\left( \mu _{0}\right) \right) \right.   \notag \\
&&\times \left( P_{-1/2+iM}^{\prime }{}^{m}\left( \mu _{0}\right)
P_{-1/2+iM}^{m}\left( -\mu _{0}\right) \right.   \notag \\
&&\left. \left. +P_{-1/2+iM}^{\prime }{}^{m}\left( -\mu _{0}\right)
P_{-1/2+iM}^{m}\left( \mu _{0}\right) \right) ^{-1}\right\} ^{1/2},
\label{eq18}
\end{eqnarray}

\noindent where $m=0,\;1,\;2,\;\ldots $ and the indices $L$, $M$, which
characterize the modes, continuously vary from 0 to the infinity; $\mu
_{0}=d/f$ ($d$ is half of the smallest distance between the hyperboloids and
$f$ is half of the distance between the hyperboloids' focuses). The
corresponding wave functions (electric potentials) in the gap ($-\mu
_{0}<\mu <\mu _{0}$) are

\begin{eqnarray}
\psi _{m}^{\left( L\right) } &=&P_{-1/2+iL}^{m}\left( \eta \right) \left(
P_{-1/2+iL}^{m}\left( \mu \right) -P_{-1/2+iL}^{m}\left( -\mu \right)
\right) ,  \notag \\
\psi _{m}^{\left( M\right) } &=&P_{-1/2+iM}^{m}\left( \eta \right) \left(
P_{-1/2+iM}^{m}\left( \mu \right) +P_{-1/2+iM}^{m}\left( -\mu \right)
\right) ,  \label{eq19}
\end{eqnarray}

\noindent where the prolate spheroid coordinates $\eta ,\;\mu $ are related
with the Cartesian coordinates by the equations

\begin{eqnarray}
\left( \eta -\mu \right) f &=&\sqrt{R^{2}+f^{2}-2fz},  \notag \\
\left( \eta +\mu \right) f &=&\sqrt{R^{2}+f^{2}+2fz},  \notag \\
R^{2} &=&x^{2}+y^{2}+z^{2}.  \label{eq20}
\end{eqnarray}

Inside the hyperboloids the expressions for wave function of plasmonic atoms
and plasmonic molecules are analogous to (\ref{eq19}).

The axis-symmetrical ($m=0$) spectra for several values of $L$,
$M$ =1, 3, 5, 10 are shown in Fig.~\ref{fig:11}b as functions of
distance between hyperbolic tips. The spatial distribution of
corresponding wavefunctions (electric potentials) for $L$, $M$ =1
are given in Fig.~\ref{fig:12}.

As seen from the figures, for both the symmetrical and antisymmetrical
plasmonic oscillations there is observed a strong localization of the modes,
and this ensures an effective interaction with the ordinary atoms, molecules
or quantum dots with  plasmonic atoms and molecules near the gap between tips.
 One more interesting fact is that the spectra of these
oscillations are quite close to the oscillations in a two-sphere cluster.
This allows one to consider the localized plasmonic oscillations in the gap
between semi-infinite bodies as the virtual plasmonic atoms and molecules.

\section{Possible Applications of plasmonic molecules}

\label{sec:4}

Obviously, strong localization of plasmonic molecules may be suitable for a
number of applications. This is particularly important when one needs to
ensure effective interaction between the nanolocalized light sources
(molecules and nanocrystal quantum dots) and the nanoparticles and
nanostructures, as well as the nano-electromechanical devices \cite{ref20},
in which the van der Waals forces play an important role \cite{ref41,ref42}.

\subsection{The effect of the plasmonic molecules on the van der Waals
forces (V.V. Klimov, A. Lambrecht to be published)}

\label{sec:41}

As is well known, the van der Waals forces are associated with spatial
dependence of the vacuum fluctuation energy density. If the distance between
the plasmonic nanoparticles is small, then the main contribution to the van
der Waals energy comes from zero (vacuum) oscillations of the plasmonic
modes, i.e. the plasmonic atoms and molecules \cite{ref44}.

In case of two identical plasmonic nanospheres the contribution to
the van der Waals energy comes from zero oscillations of the
antisymmetric and symmetric unbound plasmonic atoms (L- and
T-modes) and plasmonic molecules (M-modes)

\begin{eqnarray}
U_{vdW} &=&U_{vdW}^{M}+U_{vdW}^{L}+U_{vdW}^{T}  \notag \\
&=&\frac{{\hbar }}{{2}}\left\{ \sum\limits_{M=1}^{\infty }\omega
_{0}^{M}+\sum\limits_{L=1}^{\infty }\omega
_{0}^{L}+\sum\limits_{T=1}^{\infty }\omega _{0}^{T}\right\} \notag
\\ &&+\hbar \left\{ \sum\limits_{M,m=1}^{\infty }\omega
_{m}^{M}+\sum\limits_{L,m=1}^{\infty }\omega
_{m}^{L}+\sum\limits_{T,m=1}^{\infty }\omega _{m}^{T}\right\} .
\label{eq21}
\end{eqnarray}

In spite of the fact that the expressions for the van der Waals energy of
different modes are formally similar, the noted contributions result in
totally different physical consequences. Figure ~\ref{fig:6} shows that the
energy of antisymmetric unbound states of the plasmonic atoms grows with
distance, and this results in the particle attraction. In a similar manner,
the energy of the plasmonic molecules diminishes with distance, and thus
results in the repulsion of the nanospheres. Symmetric states of unbound
plasmonic atoms also lead to a very weak repulsion.

The attraction or repulsion conditioned by different plasmonic oscillations
is of a simple physical nature. First consider the axis-symmetrical
oscillations ($m=0$). Here the longitudinal antisymmetric modes are
conditioned by the oscillations of the parallel dipoles which are attracting
in accordance with the law of electrostatics (see Fig.~\ref{fig:14} and Fig.~%
\ref{fig:7}). In case of symmetrical oscillations the dipoles have opposite
directions, and this results in the dipole repulsion (see Fig.~\ref{fig:14}
and Fig.~\ref{fig:7}).

For $m=1$, i.e. when the wave functions are described by the law of $\cos
\varphi $ or $\sin \varphi $ ($\varphi $ is the angle describing the
rotation around symmetry axis), the directions of the dipoles corresponding
to different plasmonic states are changed. For antisymmetric modes the
dipole moments in different spheres should be crosswise to the symmetry axis
and should have opposite orientations (Fig.~\ref{fig:14} and Fig.~\ref{fig:7}%
). It is easily seen that this results again in the effect of attraction of
the unbound plasmonic atoms. For the plasmonic molecules the dipole momenta
should also be crosswise to the symmetry axis, but, due to the symmetry with
respect to z coordinate, should have the same direction (see Fig.~\ref%
{fig:14}). This again leads to the repulsion of nanospheres with plasmonic
molecules.

Fig.~\ref{fig:15} illustrates the dependence of the contributions
from different plasmonic states into the van der Waals energy on
the distance between the spheres. This was obtained by direct
summation of all the plasmonic modes in (\ref{eq21}). As it was
clear beforehand, the symmetric modes lead to the repulsion, while
the longitudinal antisymmetric modes lead to the attraction of the
nanoparticles. Unexpectedly, the repulsion contribution from
plasmonic molecules is almost equal to the attraction contribution
of the unbound plasmonic atoms. As a result, the total van der
Waals energy grows with distance, i.e. it is the attraction
energy, but the amount of this energy is by an order smaller than
the energy to be obtained without taking into account the
contribution from the plasmonic molecules.

Direct measurement of the van der Waals force between the plasmonic
nanoparticles may form the basis of the experimental proof of the plasmonic
molecule existence.

\subsection{The effect of plasmonic molecules on the emission of ordinary
molecules and quantum dots}

\label{sec:42}

Since the plasmonic molecules particularly effectively interact with highly
inhomogeneous electric fields, then one can develop different detectors of
the ordinary single molecules or quantum dots. Fig.~\ref{fig:16} illustrates
the dependence of the radiation decay rate of spontaneous emission of
different molecules, which fall within the gap between the two spheres, on
the radiation wavelength. It is assumed that the nanospheres are made of Na
with the plasmonic resonances in the visible range (Fig.~\ref{fig:16}a) or
SiC with the phonon-polariton resonances in the IR range (Fig.~\ref{fig:16}%
b).  The peaks in the right-hand part of the figures correspond to the
interaction of an ordinary molecule and the unbound plasmonic atoms, while
the peaks in the left-hand part correspond to the interaction with the
plasmonic molecules.

The analysis of the figures shows that, similar to the case of the van der
Waals energy, the interaction with plasmonic molecules is of greater
importance than that with the symmetric unbound plasmonic atoms. Of
fundamental importance is the fact that in a homogeneous external field the
plasmonic molecules are not effectively excited. This results in the
following: the clusters of two and more nanoparticles (where the geometry
allows one to excite the plasmonic molecules of a certain frequency), can be
used in the development of effective nanodetectors of single molecules of
high signal/noise ratio, since the external fields at the plasmonic molecule
frequency fail to excite the plasmonic molecule effectively.

On the other hand, the effective interaction between the plasmonic and
ordinary atoms and molecules can be applied both to the development of
SPASER and single atom nanolaser, and the sets of such devices.

\subsection{Large cluster or infinite array of nanoparticles}

\label{sec:43}

Probably, the most interesting consequence of the plasmonic
molecule strong localization near the gap between the
nanoparticles of arbitrary shapes lies in possible existence of
plasmonic molecules in the clusters with more than two particles.
Figure ~\ref{fig:13} illustrates a cluster of nine metallic
nanoparticles, in which different configurations of plasmonic
molecules may exist. In each of the twelve gaps a plasmonic
molecule having the properties defined by the curvature radii of
the closest surfaces and given by expressions
(\ref{eq9})-(\ref{eq11}) may be excited. In this system one can
also excite the pairs, triples, quadruples, etc., of plasmonic
molecules. Under simultaneous excitation of several plasmonic
molecules a weak interaction between the molecules may be easily
taken into account using the perturbation theory. The account for
such an interaction results in the hybridization of the wave
functions of single plasmonic molecules and a corresponding
splitting of the spectrum. Similar reasoning is also fully
applicable to the large-size arrays of nanoparticles, and this
allows one to speak about plasmonic molecular crystals. In other
words, a strong spatial localization of plasmonic molecules and
their weak interaction with each other allows an almost linear
scaling of any simpler structure of plasmonic molecules. To put it
in simpler words, the plasmonic molecules allow one, similar to
the Lego bricks, to produce the optoelectronic nanodevices of any
complexity.

In principle, the excitation of plasmonic molecules in the clusters and
arrays may be performed with the help of homogeneous optical fields of a
corresponding resonant frequency. However, such a method of excitation is
not highly effective due to weak polarizability of plasmonic molecules. A
positive feature of such inefficiency is a weak sensitivity of plasmonic
molecules to the external noise fields. A more effective method to excite
the plasmonic molecules lies in the excitation by means of spontaneous or
stimulated radiation (SPASER \cite{ref12})of usual molecules at a frequency
of plasmonic molecule oscillation . In this case, depending on the
orientation and location of the ordinary molecules ( or nanocrystal quantum
dots) with respect to the exciting field, the plasmonic molecules may be
produced in the given part of a cluster or array, and, thus, to produce
different initial conditions for the data processing and transfer.

\section{Conclusion}

\label{concl}

In this work the analysis has been made of the plasmonic oscillations in
single nanoparticles and the clusters of nanoparticles. It has been shown
that all the plasmonic oscillations may be classed by the degree of their
localization. Such oscillations where the localization is mostly defined by
the size of single nanoparticles we propose to name "the plasmonic atoms",
and those strongly localized in the nanoparticle contact zones (without
electrical contact) we propose to name "the plasmonic molecules'', since
they are the bound states of plasmonic atoms. It has been demonstrated that
the plasmonic molecules almost do not interact between each other, and, so,
can form the structures of arbitrary spatial complexity. The methods of
plasmonic molecule excitation and its applications in the description of
NEMS and the nanodetectors of single molecules are briefly discussed as well.

\begin{acknowledgments} The authors are grateful to the Russian
Foundation for Basic Research (grants 05-02-19647 and 07-02-01328)
and the Russian Academy of Sciences and Centre de Recherche
Scientifique (France) for financial support of this work.

One of the authors (D.V.G.) is thankful to the Young scientists
support Program of the Educational-Scientific Center of P.N.
Lebedev Physical Institute and the RAS Presidium Program "Support
of young scientists" for a partial financial support of the work.
\end{acknowledgments}

\pagebreak
\newpage

\begin{center}
\bigskip {\LARGE List of Figure Captions}
\end{center}

\bigskip

\begin{figure}
\centering \includegraphics[height=9cm,angle=0]{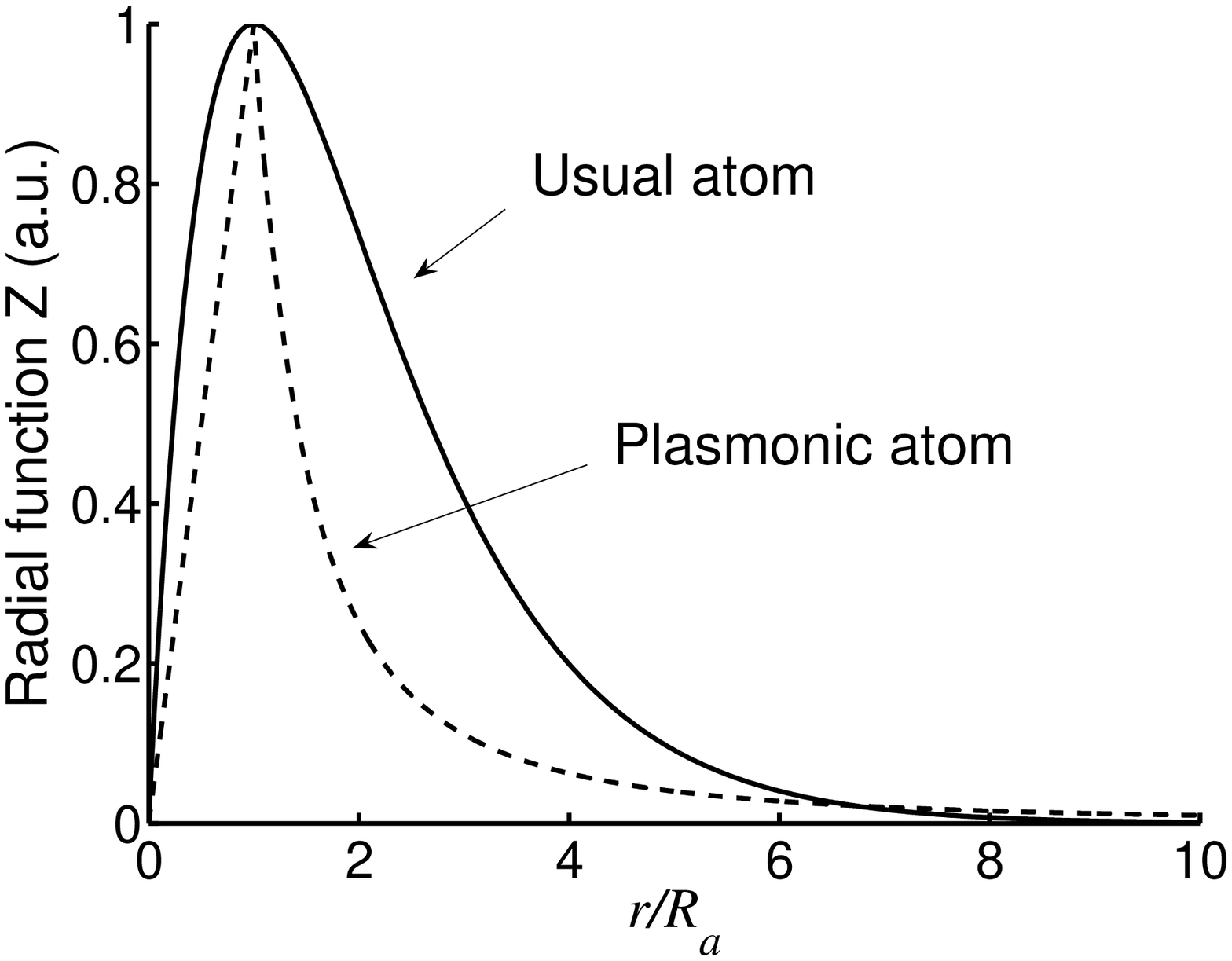}
\caption{} \label{fig:1}
\end{figure}

\textbf{Figure ~\ref{fig:1}} Radial wave functions (a.u.) of usual
and plasmonic atoms. The parameter $R_{a} = \sqrt{2}r_{B}$ or
$R_{a} = R_{0}$. The solid line corresponds to an ordinary atom
and the dotted line, to the plasmonic atom.

\begin{figure}
\centering \includegraphics[height=9cm,angle=0]{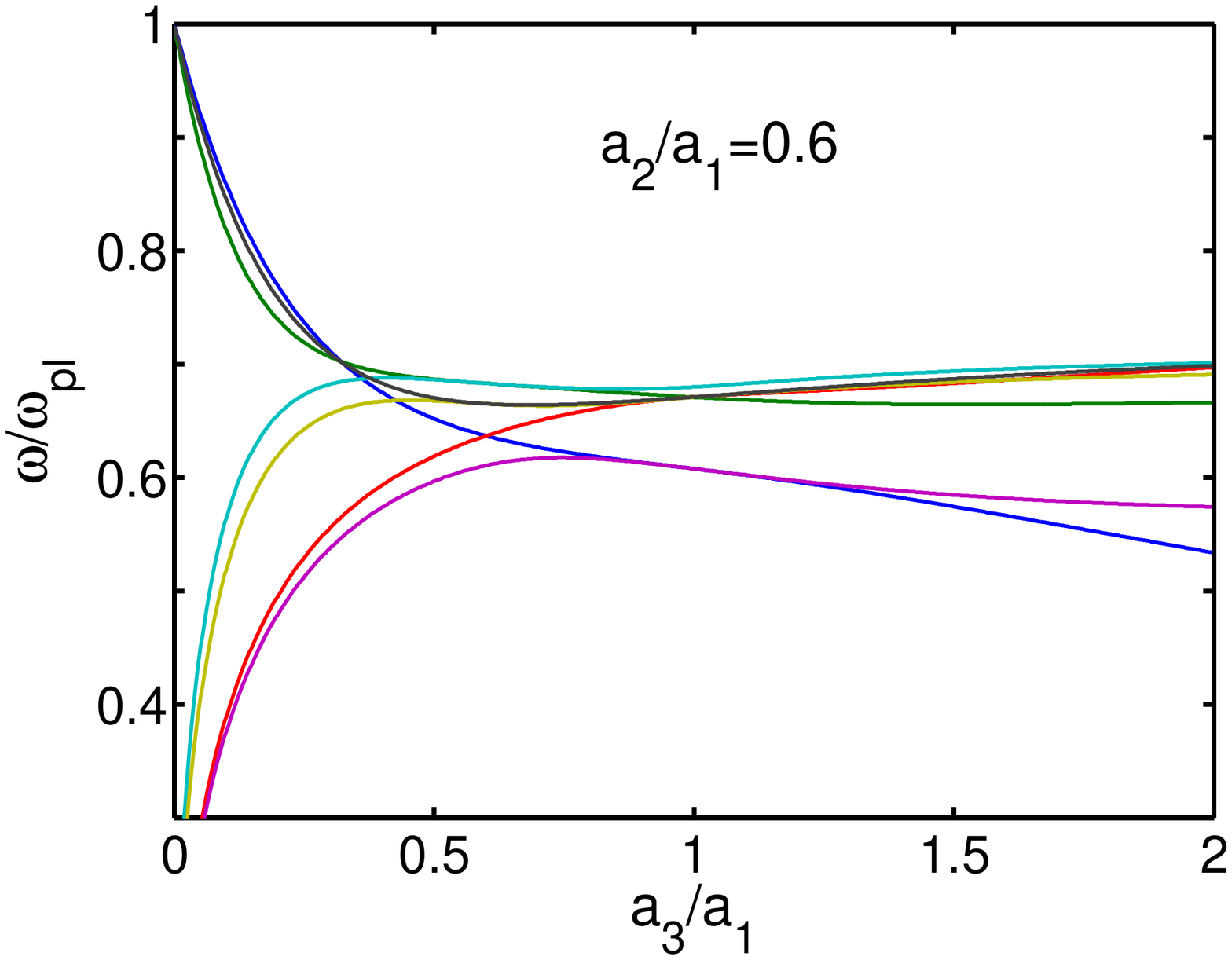}
\caption{} \label{fig:2}
\end{figure}

\textbf{Figure ~\ref{fig:2}} Plasmon frequencies for triaxial
nanoellipsoid for $n=3$ as a function of aspect ratio $\
a_{3}/a_{1}$ for $a_{2}/a_{1}$ = 0.6. Different colors denote
different values $m$=1, 2, \ldots\ 7.

\begin{figure}
\centering \includegraphics[height=5cm,angle=0]{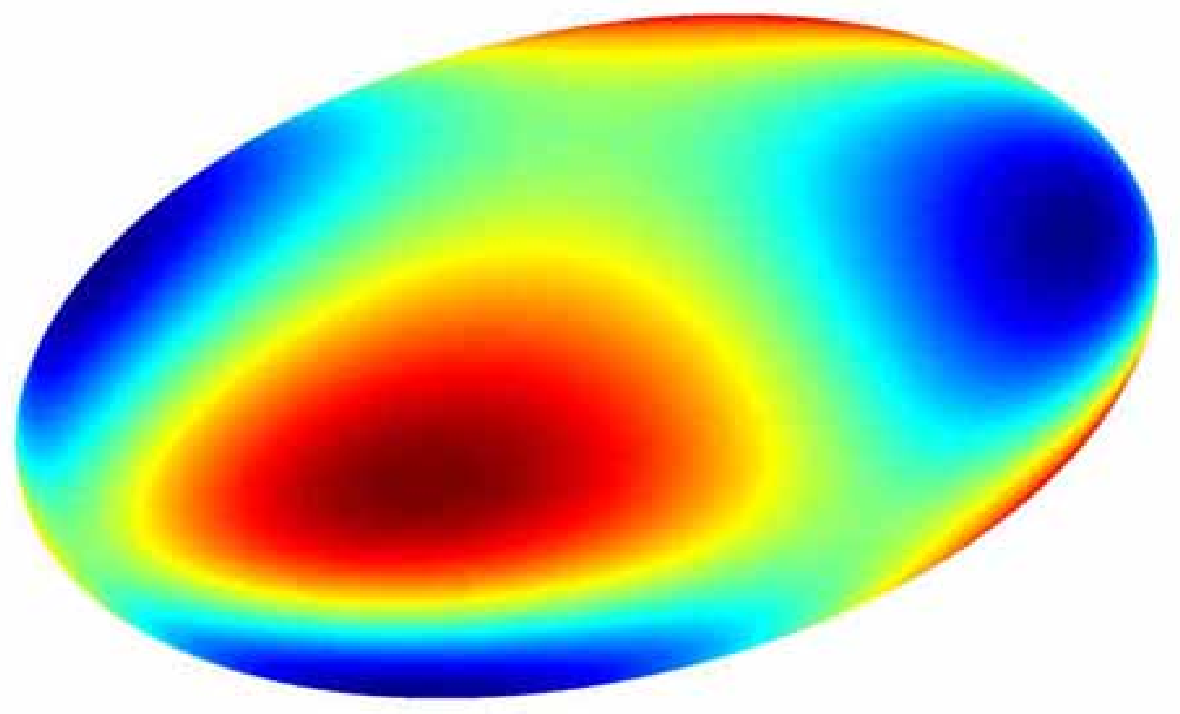}
\caption{} \label{fig:3}
\end{figure}

\textbf{Figure ~\ref{fig:3}} Surface charge distribution (a. u.)
for a plasmonic atom in triaxial metallic nanoellipsoid for $n=3$
and $m=7$. The ratios of ellipsoid semi-axes $a_{2}/a_{1}$ = 0.7
and $a_{3}/a_{1}$ = 0.4.

\begin{figure}
\centering \includegraphics[height=9cm,angle=0]{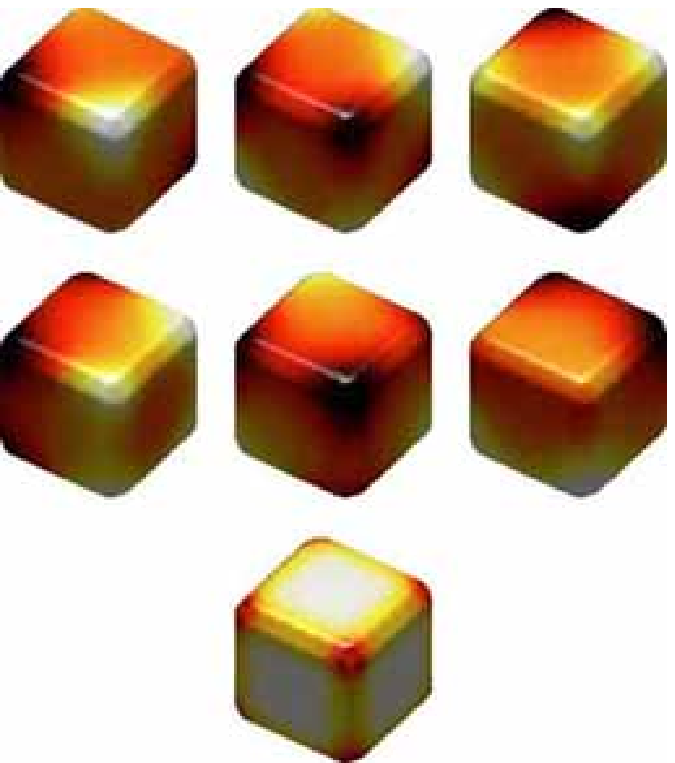}
\caption{} \label{fig:4}
\end{figure}

\textbf{Figure ~\ref{fig:4}} Normal modes (a.u.) corresponding to
the six major absorption peaks of a cube \protect\cite{ref27}.

\begin{figure}
\centering \includegraphics[height=10cm,angle=0]{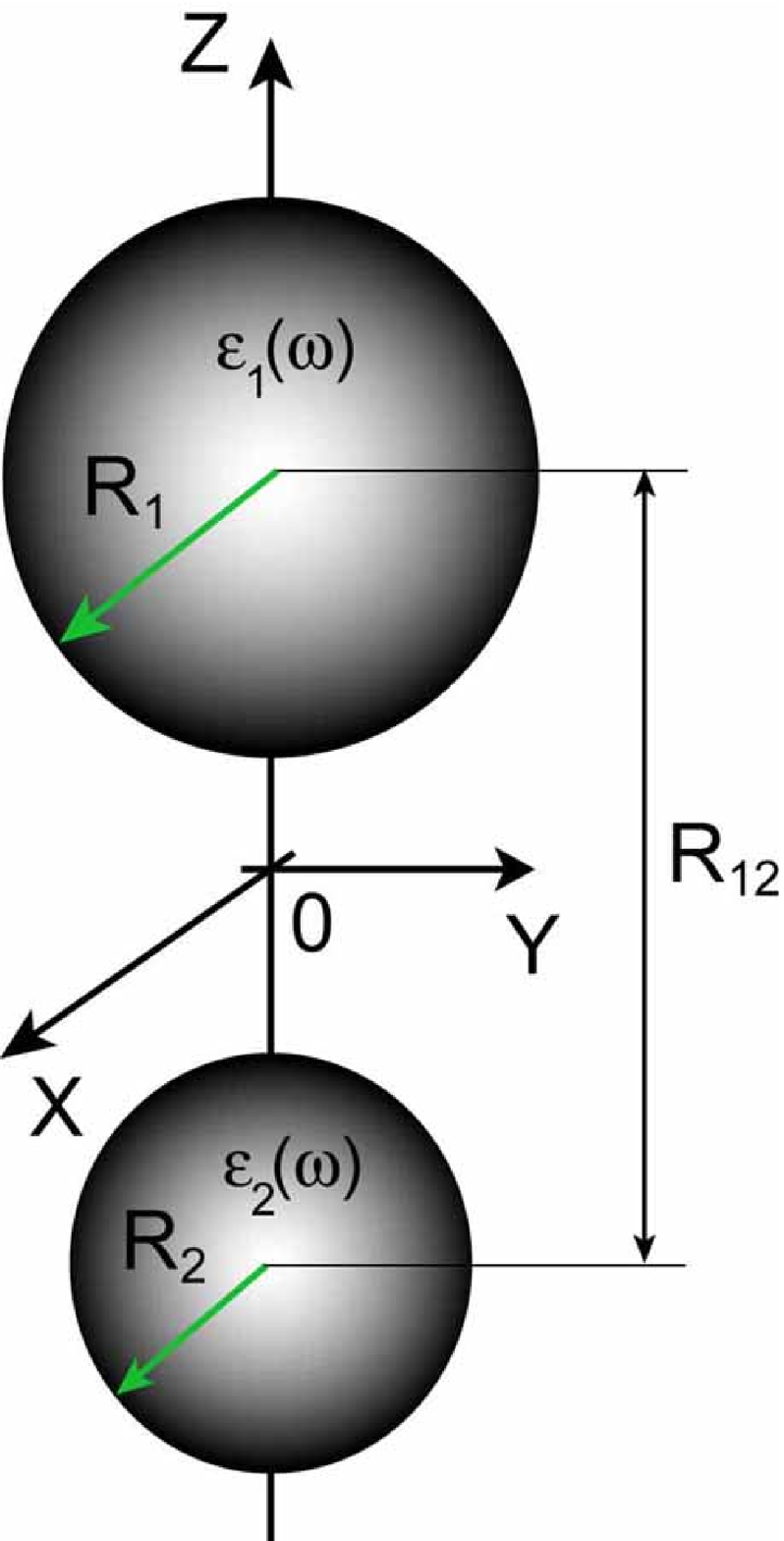}
\caption{} \label{fig:5}
\end{figure}

\textbf{Figure ~\ref{fig:5}} Geometry of a two-sphere cluster.

\begin{figure}
\centering \includegraphics[height=9cm,angle=0]{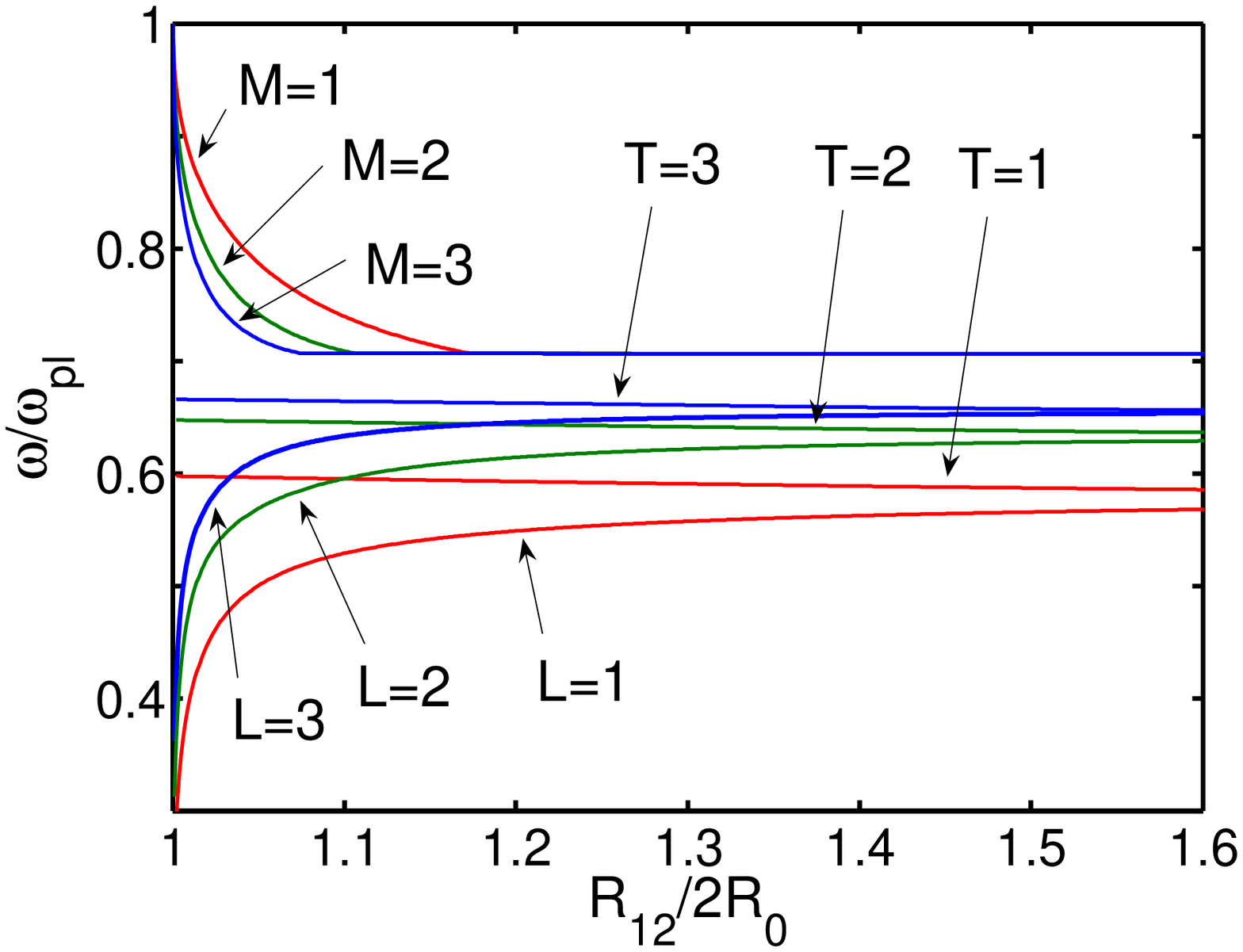}
\caption{} \label{fig:6}
\end{figure}

\textbf{Figure ~\ref{fig:6}} Spectrum of plasmon oscillations in a
two-sphere cluster for $m=1$ for different modes L, M, T = 1, 2, 3
as a function of distance between the spheres.

\begin{figure}
\centering \includegraphics[height=11cm,angle=0]{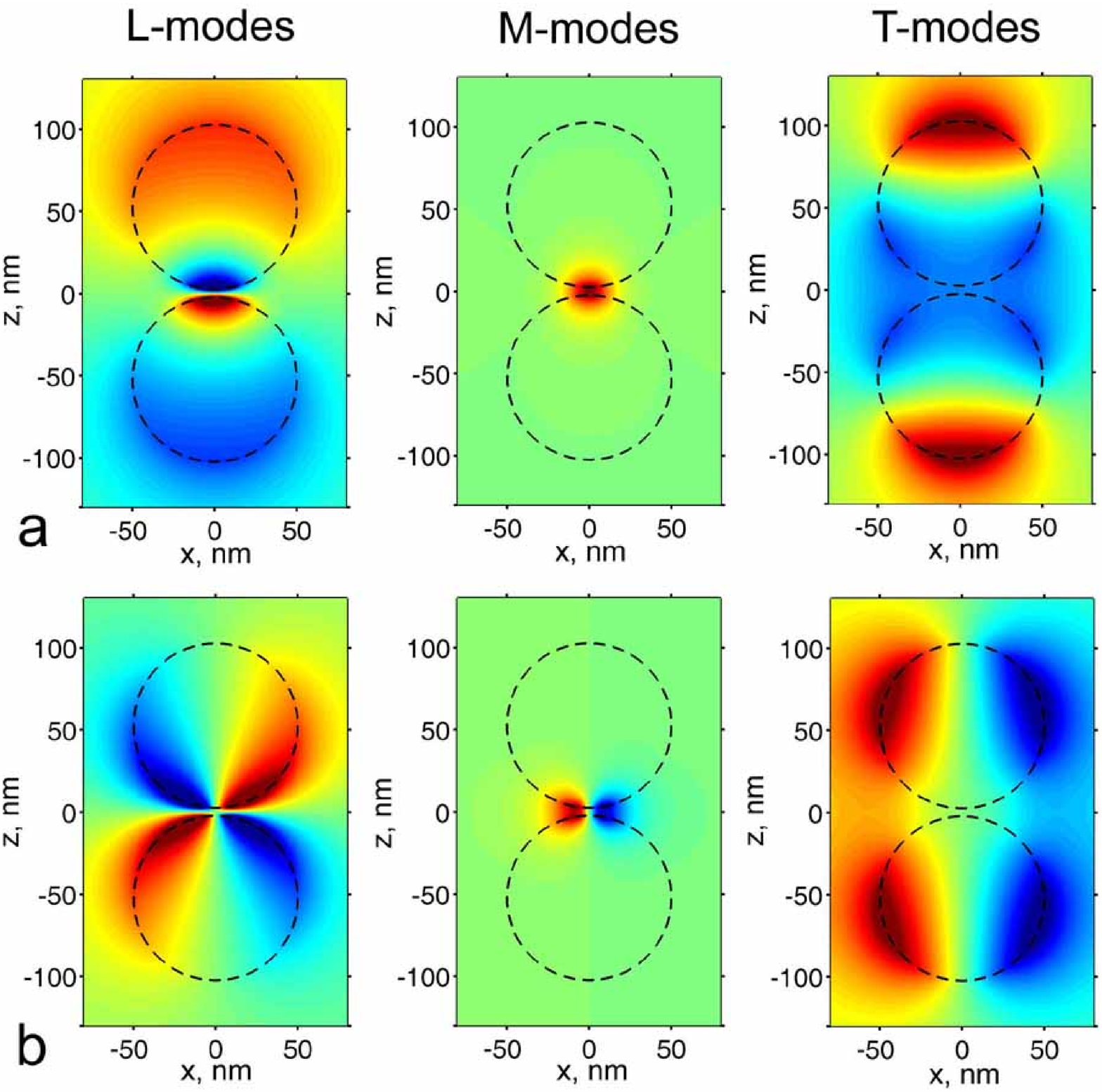}
\caption{} \label{fig:7}
\end{figure}

\textbf{Figure ~\ref{fig:7}} Spatial distribution of electric
potential of the modes with L, M, T = 1  in $x$-$z$ plane. a, Case
$m=0$. b, Case $m=1$. Radii of the spheres, $R_{0}$=50 nm;
distance between their centers, $R_{12}$=105 nm. Surface of the
nanospheres is denoted by a dotted line.

\begin{figure}
\centering \includegraphics[height=8cm,angle=0]{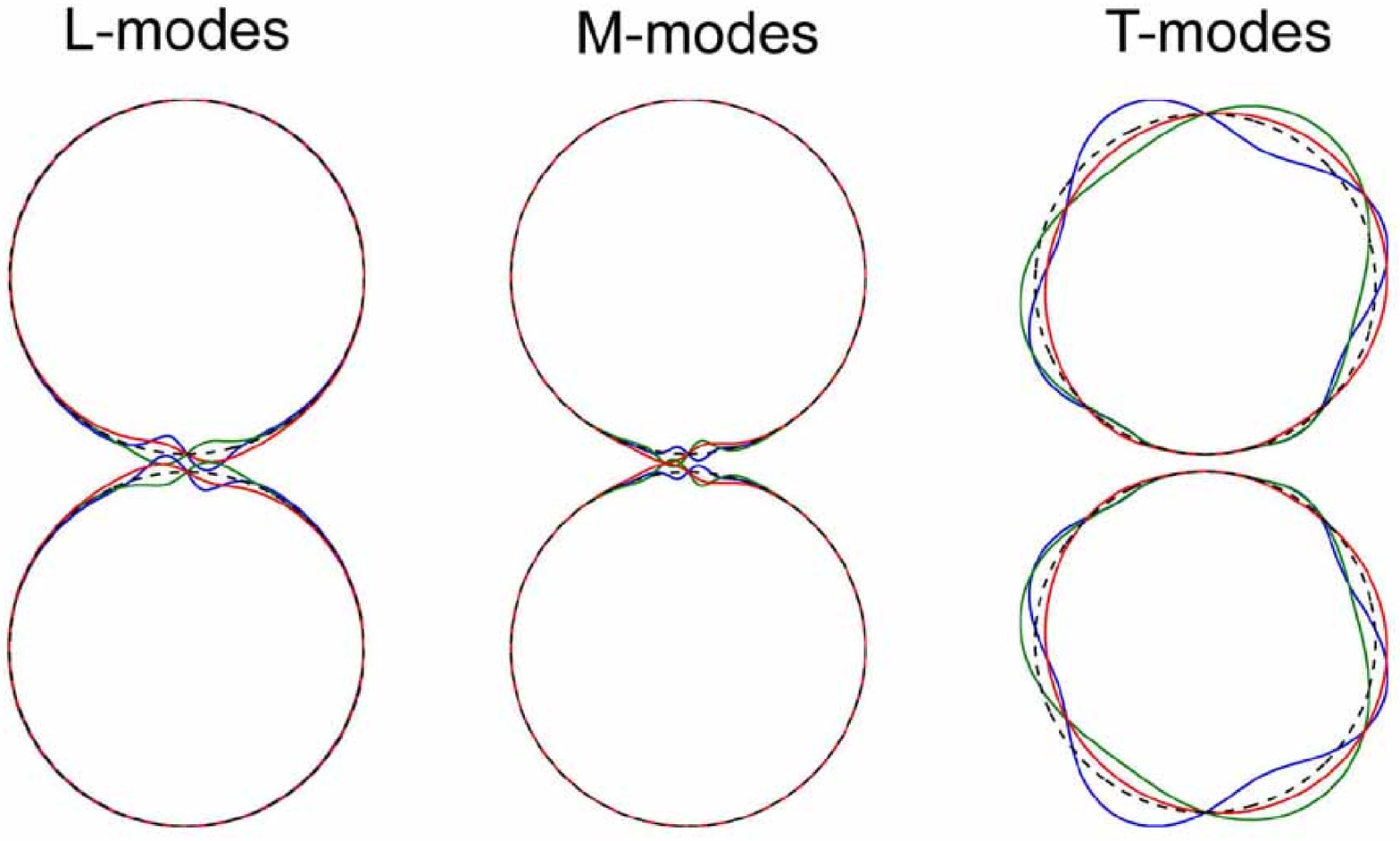}
\caption{} \label{fig:8}
\end{figure}

\textbf{Figure ~\ref{fig:8}} Surface charge (a.u.) of different
modes with L, M, T= 1, 2, 3 at $m=1$. Radii of the spheres
$R_{0}$=50 nm; distance between their centers, $R_{12}$=105 nm.
Surface of the nanospheres is denoted by the dotted line.

\begin{figure}
\centering \includegraphics[height=6cm,angle=0]{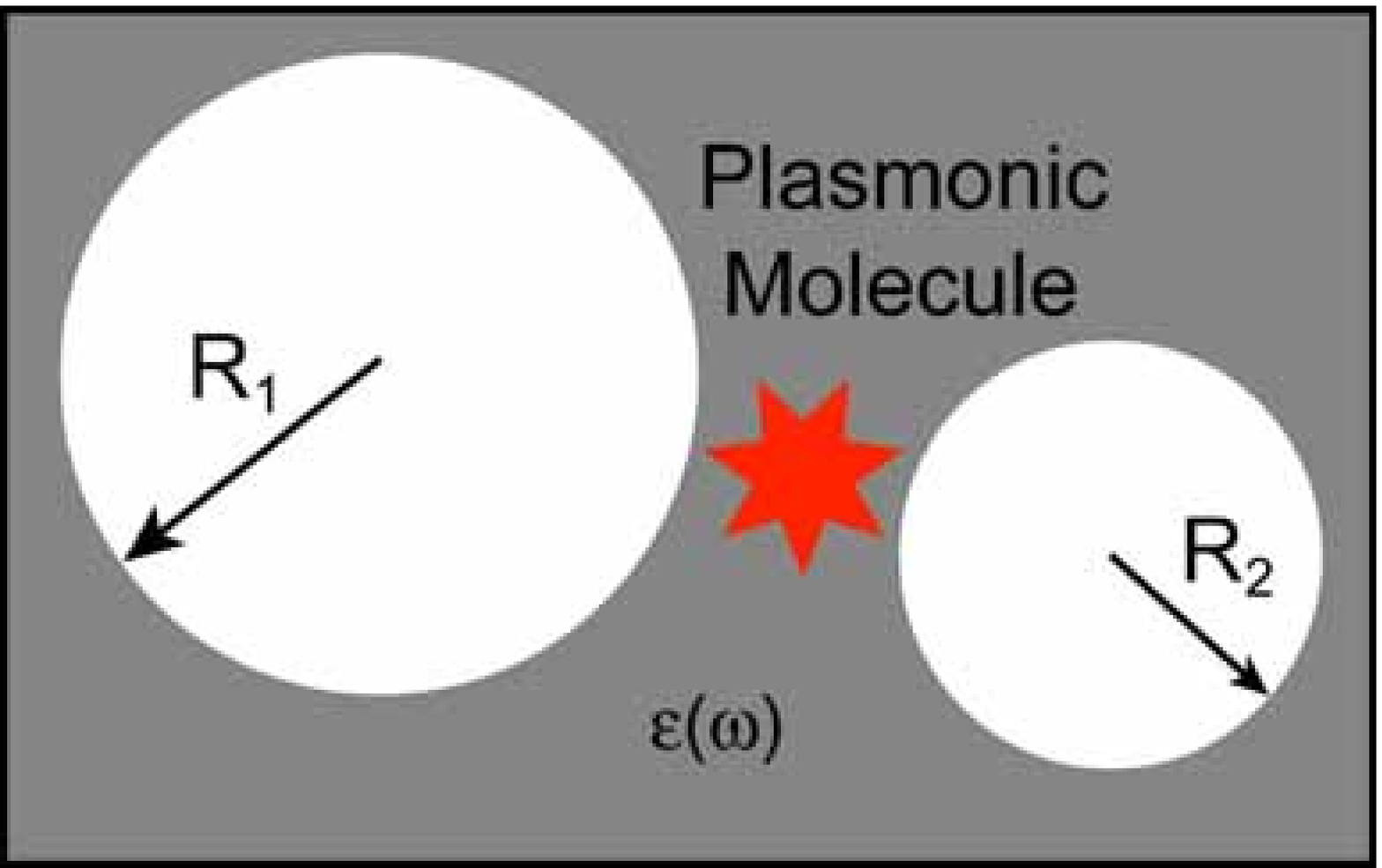}
\caption{} \label{fig:9}
\end{figure}

\textbf{Figure ~\ref{fig:9}} Plasmonic molecules in the gap
between nanobubbles in metal.

\begin{figure}
\centering \includegraphics[height=7cm,angle=0]{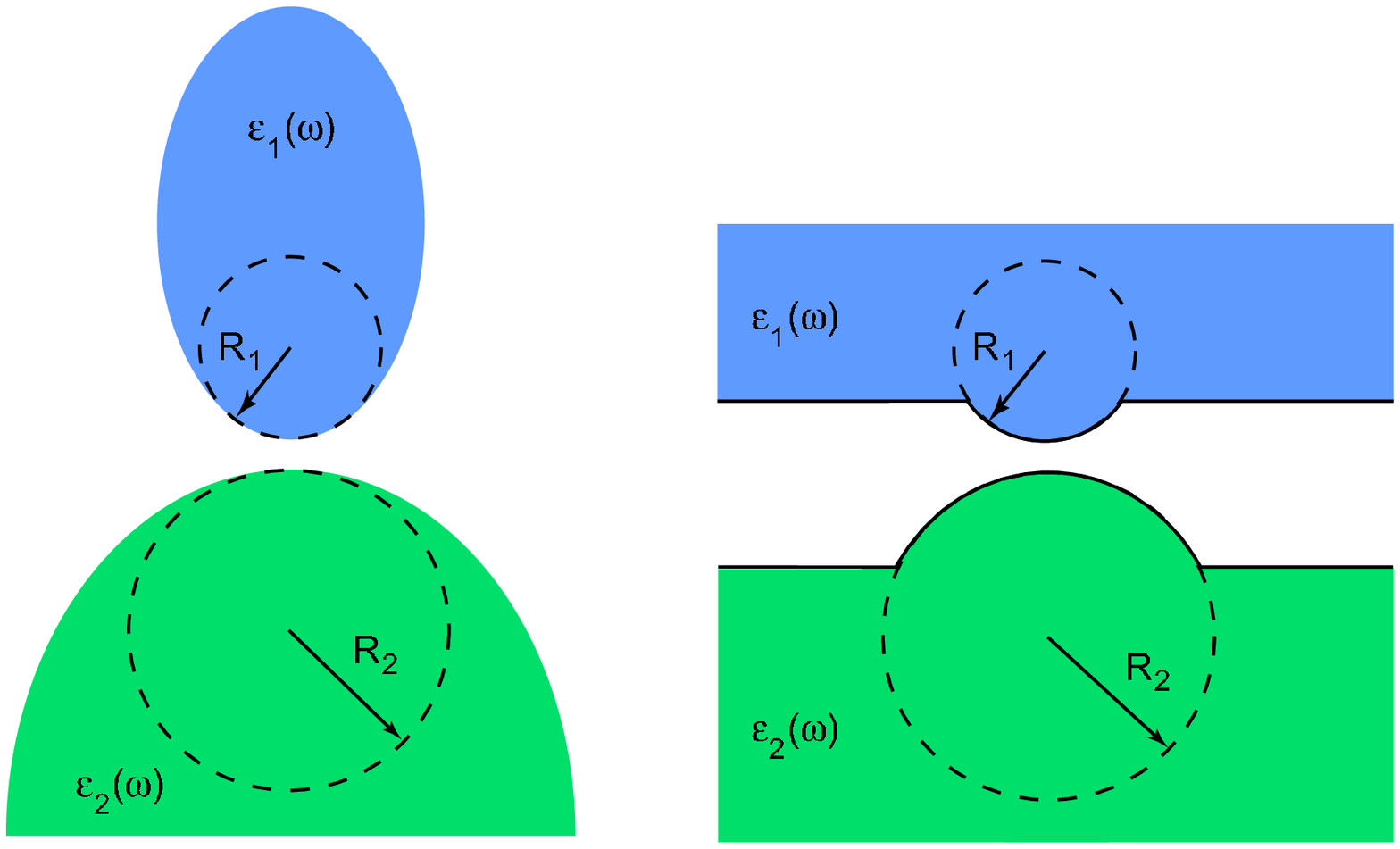}
\caption{} \label{fig:10}
\end{figure}

\textbf{Figure ~\ref{fig:10}} Approximation of nonspherical
nanoparticles with equivalent spherical ones,which allows to
estimate properties of plasmonic molecules.

\begin{figure}
\centering \includegraphics[height=18cm,angle=0]{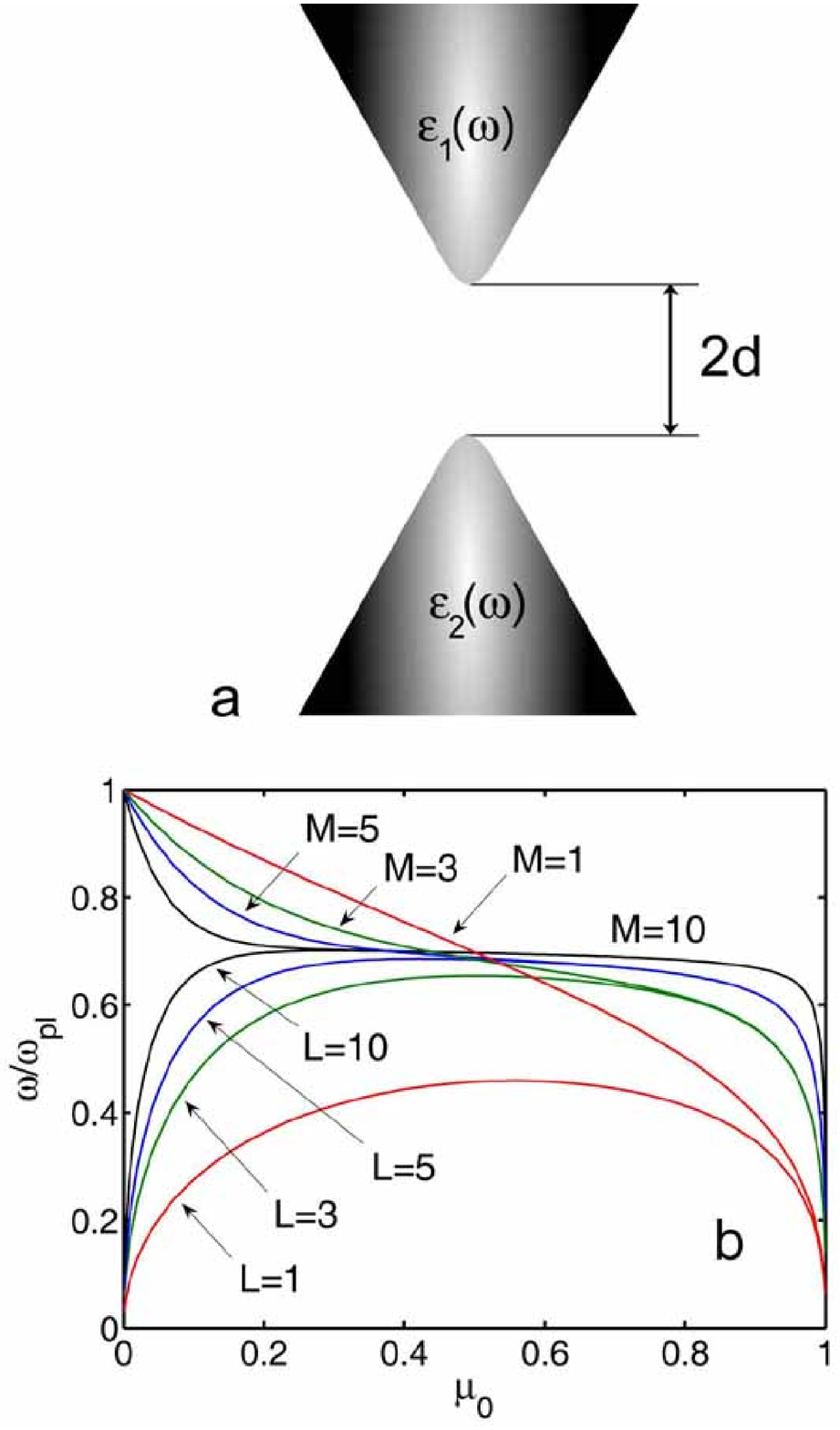}
\caption{} \label{fig:11}
\end{figure}

\textbf{Figure ~\ref{fig:11}} Plasmonic molecules in the gap
between two hyperboloids. a, Geometry of the problem. b, Spectra
for $m=0$ and $L$, $M$ = 1, 3, 5, 10.

\begin{figure}
\centering \includegraphics[height=9cm,angle=0]{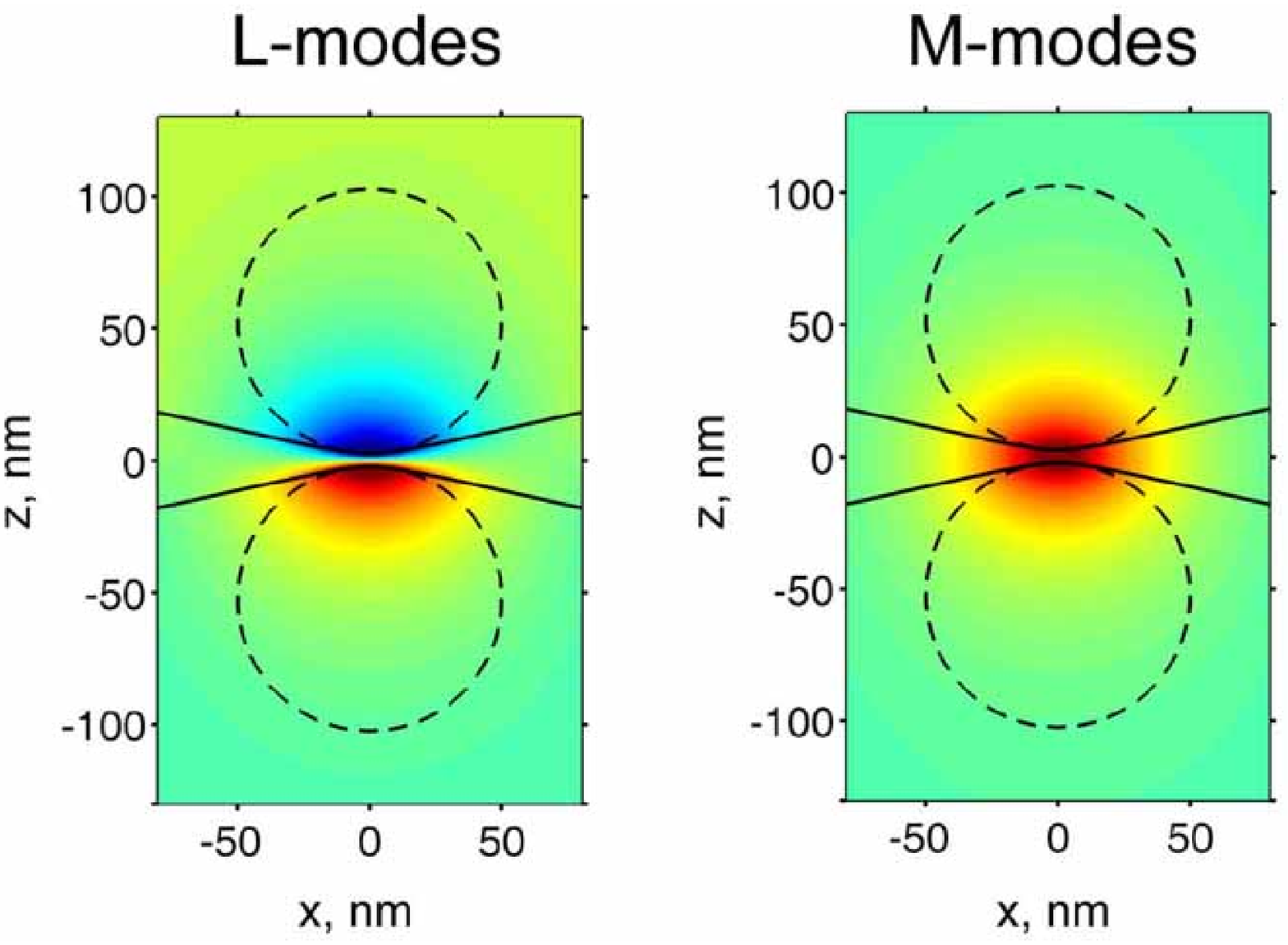}
\caption{} \label{fig:12}
\end{figure}

\textbf{Figure ~\ref{fig:12}} Spatial distribution of wave
functions for localized plasmonic oscillations in the gap between
two hyperboloids for $m=0$ and $L,$ $M$ = 1. Distance between
hyperboloids, 2$d$=5 nm; the least curvature radius is 50 nm (cf.
Fig.~\protect\ref{fig:7}a). Solid line depicts surfaces of
hyperboloids; dotted line corresponds to the cluster of two
nanospheres with $R_{0}$ = 50 nm and $R_{12}$ = 105 nm.

\begin{figure}
\centering \includegraphics[height=9cm,angle=0]{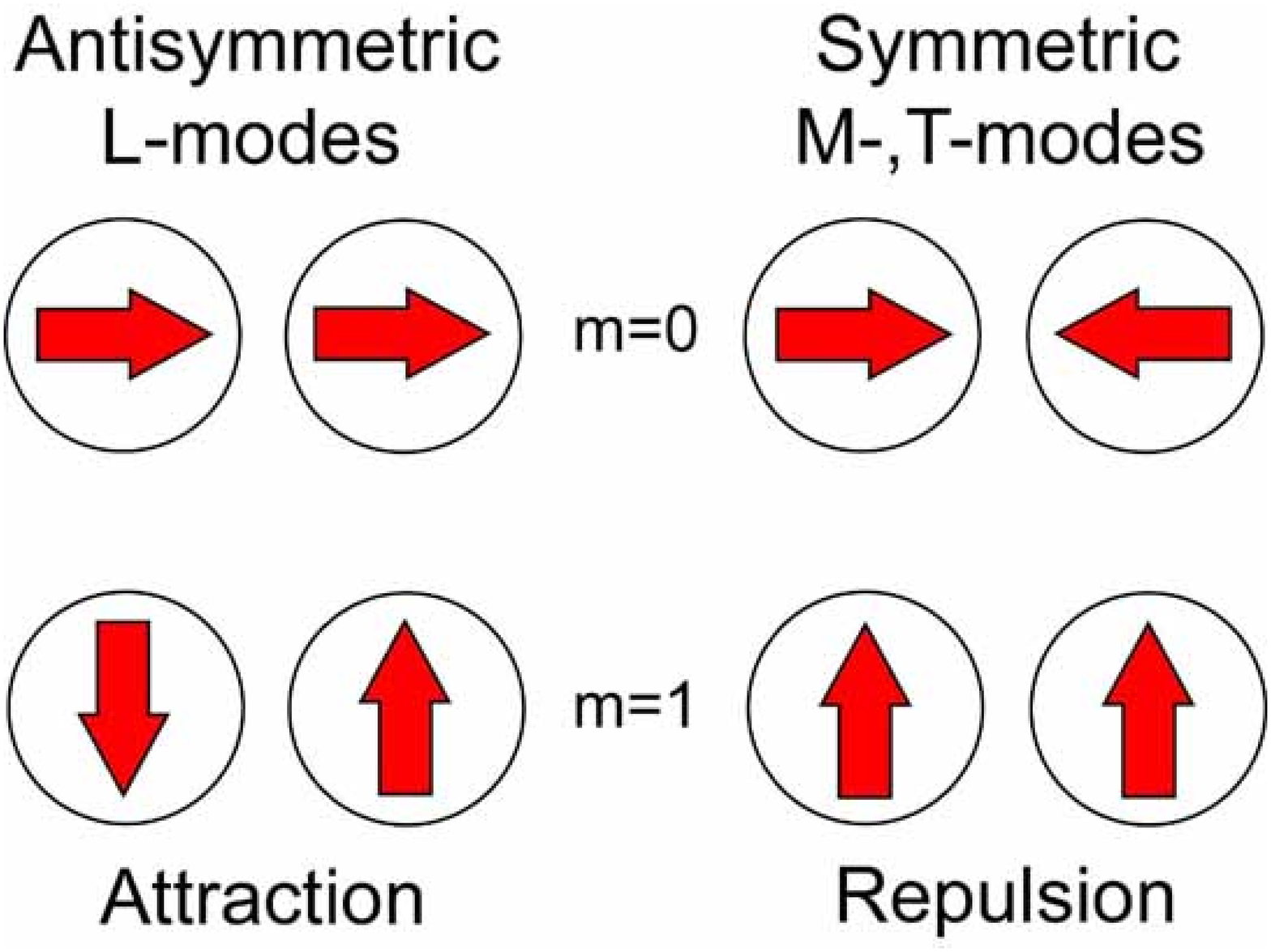}
\caption{} \label{fig:14}
\end{figure}

\textbf{Figure ~\ref{fig:14}} Illustration of attraction and
repulsion of the nanospheres due to different plasmonic states.

\begin{figure}
\centering \includegraphics[height=9cm,angle=0]{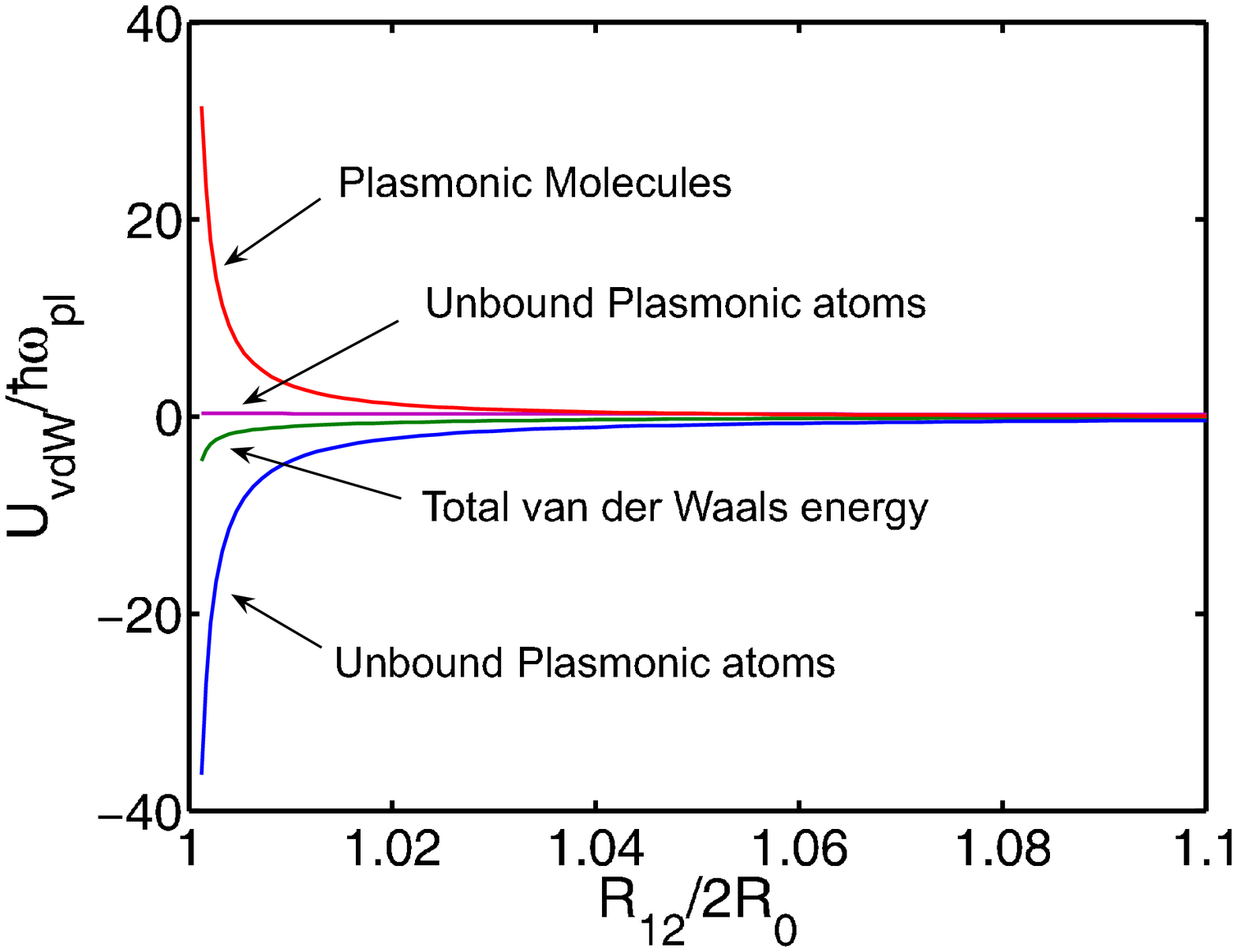}
\caption{} \label{fig:15}
\end{figure}

\textbf{Figure ~\ref{fig:15}} Van der Waals energy of two-sphere
cluster defined by different plasmonic states versus distance
between the spheres.

\begin{figure}
\centering \includegraphics[height=18cm,angle=0]{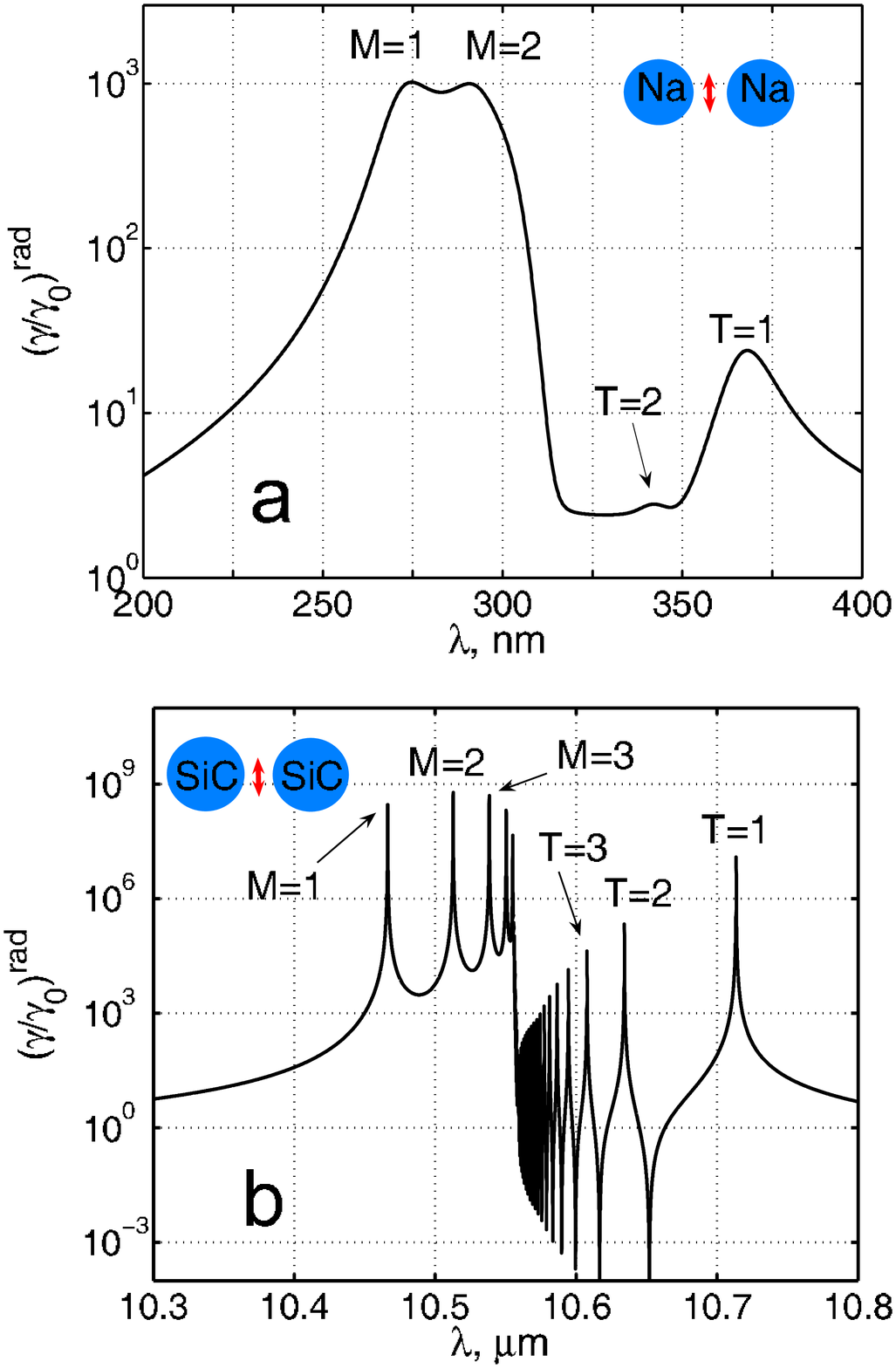}
\caption{} \label{fig:16}
\end{figure}

\textbf{Figure ~\ref{fig:16}} Relative radiative decay rate of a
molecule located between two nanospheres versus the radiation
wavelength. a, Nanospheres made of sodium (Na). b, Nanospheres
made of silica-carbone (SiC). In both cases, $R_{0}$ = 50 nm and
$R_{12}$ = 105 nm. Direction of the molecular dipole moment is
shown by arrow. The dielectric constant for Na is borrowed from
\cite{ref45}, for SiC, from \cite{ref46}.

\begin{figure}
\centering \includegraphics[height=7cm,angle=0]{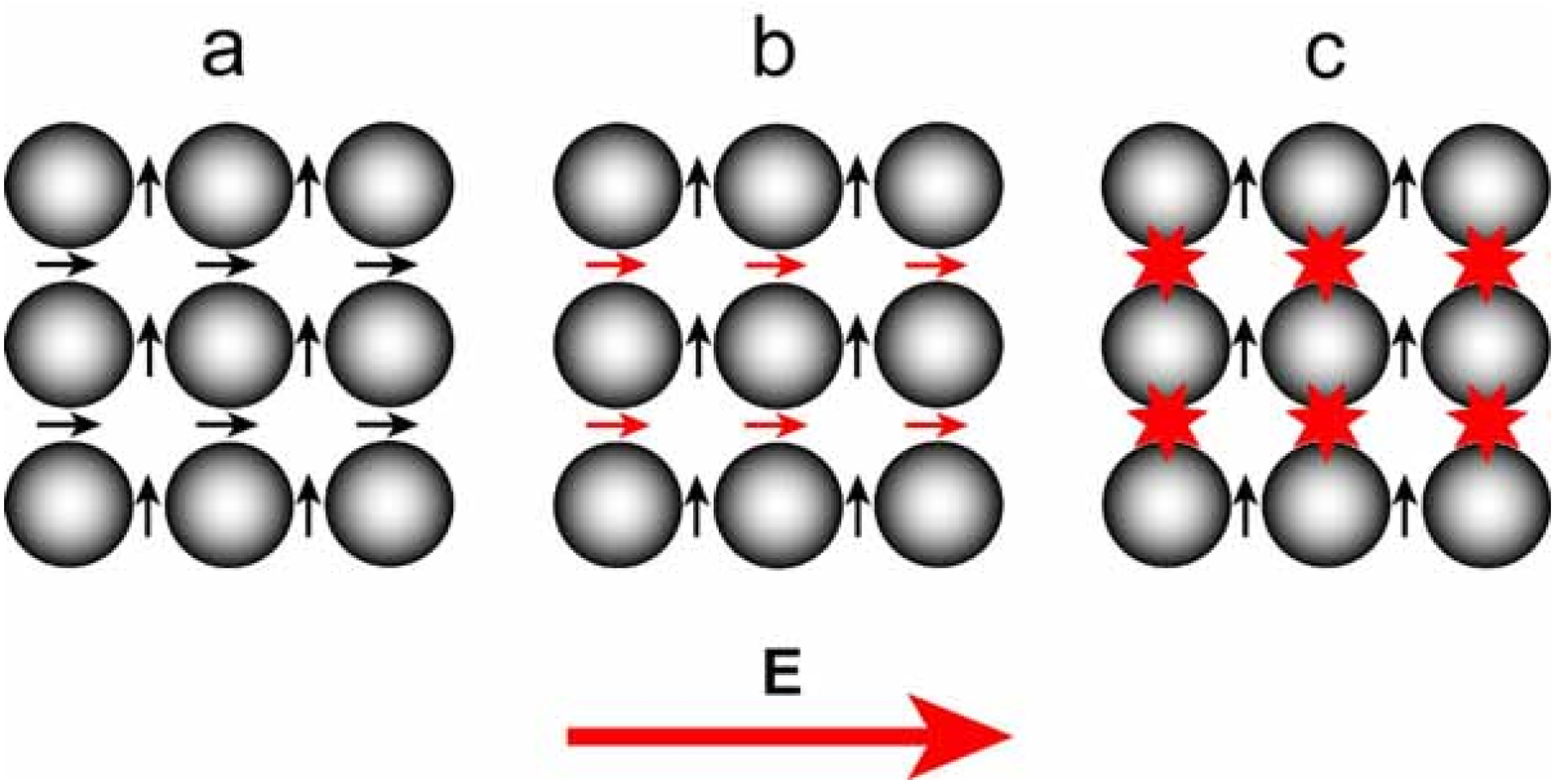}
\caption{} \label{fig:13}
\end{figure}

\textbf{Figure ~\ref{fig:13}} Excitation of plasmonic molecules in
clusters of nine nanospheres with usual molecules. a, Usual and
plasmonic molecules are not excited. b, External field with
horizontal polarization excite the usual molecules. c, Radiation
of usual molecules excite the plasmonic molecules with horizontal
polarization.

\newpage

\end{document}